\documentclass{article}

\usepackage{PRIMEarxiv}

\usepackage[utf8]{inputenc} 
\usepackage[T1]{fontenc}    
\usepackage{hyperref}       
\usepackage{url}            
\usepackage{booktabs}       
\usepackage{amsfonts}       
\usepackage{nicefrac}       
\usepackage{microtype}      
\usepackage{lipsum}
\usepackage{epigraph} 
\usepackage{dirtytalk}
\usepackage{csquotes}
\usepackage{amsmath}
\usepackage{fancyhdr}       
\usepackage{graphicx}       
\graphicspath{{media/}}     
\graphicspath{{Images/}}

\makeatletter
\newenvironment{chapquote}[2][2em]
  {\setlength{\@tempdima}{#1}%
   \def\chapquote@author{#2}%
   \parshape 1 \@tempdima \dimexpr\textwidth-2\@tempdima\relax%
   \itshape}
  {\par\normalfont\hfill--\ \chapquote@author\hspace*{\@tempdima}\par\bigskip}
\makeatother

\title{Causality between Sentiment and Cryptocurrency Prices
\thanks{\textit{\underline{email}} : 
lubdhak.mondal@students.iiserpune.ac.in} 
}

\author{
  Lubdhak Mondal* \\
  Department of Mathematics\\
  IISER Pune \\
 \\
   \And
  Udeshya Raj \\
  Department of Metallurgy\\
  IIT Kharagpur \\
     \And
  Abinandhan S \\
  Department of Biotechnology\\
  IIT Kharagpur \\
    \And
  Began Gowsik S\\
  Department of Humanities and Social Sciences\\
  IIT Kharagpur \\
    \And
  Sarwesh P\\
  Department of Electronics \& Electrical Communication Engineering\\
  IIT Kharagpur \\
    \And
  Abhijeet Chandra\\
  Vinod Gupta School of Management\\
  IIT Kharagpur \\
}

\begin{document}
\maketitle

\begin{abstract}
This study investigates the relationship between narratives conveyed through microblogging platforms, namely Twitter, and the value of crypto assets. Our study provides a unique technique to build narratives about cryptocurrency by combining topic modelling of short texts with sentiment analysis. First, we used an unsupervised machine learning algorithm to discover the latent topics within the massive and noisy textual data from Twitter, and then we revealed 4-5 cryptocurrency-related narratives, including financial investment, technological advancement related to crypto, financial and political regulations, crypto assets, and media coverage. In a number of situations, we noticed a strong link between our narratives and crypto prices. Our work connects the most recent innovation in economics, Narrative Economics, to a new area of study that combines topic modelling and sentiment analysis to relate consumer behaviour to narratives.
\end{abstract}

\keywords{Narrative Economics \and Twitter \and NLP \and Cryptocurrencies \and Sentiment analysis \and Short Text Clustering \and GSDMM \and BERTweet}

\section{Introduction}
\begin{chapquote}{Robert J. Shiller}
``We need to incorporate the contagion of narratives into economic theory. Otherwise, we remain blind to a very real, very palpable, very important mechanism for economic change, as well as a crucial element for economic forecasting. If we do not understand the epidemics of popular narratives, we do not fully understand changes in the economy and in economic behaviour.''
\end{chapquote}

We discovered in \cite{shiller2017narrative}, \cite{akerlof2016bread} that there is mounting evidence that narratives influence economic activity. We have come a long way from the era of handwritten news pamphlets to the digital age, and in recent years it has become abundantly clear that people are shifting away from longer formats of reading, such as news articles, and towards shorter and more concise texts, which drives the massive growth of microblogging platforms such as Twitter, Google+, Facebook, etc.

According to Shiller in \cite{shiller2017narrative} "\textit{new contagious narratives cause economic events, and economic events cause changed narratives,}" and in this paper, we attempted to connect the extraordinary rise and fall of the cryptocurrency market, which rose by nearly 1,700\% from the end of 2018 to the end of 2021 and then fell by 68\% within a year. Many investors have said that Crypto Currency is a scam and a fad , yet a huge number of new-age investors are very excited about Crypto. Nonetheless, the technological developments that have resulted from this, such as Blockchain, have sparked tremendous interest.

The major goal of this article is to identify the prevailing narratives in the cryptocurrency market utilizing brief texts from the microblogging network Twitter. With the growing popularity of these platforms, clustering short texts has become an increasingly vital job. Topic clustering in short texts, on the other hand, is difficult owing to their sparse, high-dimensional, and high-volume properties.
To measure the spread of the dominant narratives, we use an unsupervised machine-learning algorithm to the collected tweets including terms linked to cryptocurrency. The technique is unsupervised in that it infers the themes of a collection of documents without labelling the tweets or training the model prior to classifying the tweets. In accordance with Gavaldón \cite{dempster1977maximum}, with the aid of this algorithm, we reveal a handful of separate narratives certain time points where we discovered a structural break in the daily bitcoin price series beginning from January 2014 to May 2021.
The remainder of the paper is structured as follows: 

\textbf{Section} \ref{sec:Data Collection} covers how we acquired the data for this study and the findings from the structural break analysis, which helps us determine the time instance when we must collect the data.

\textbf{Section} \ref{sec:Methodologies} describes our research method, which consists of four components. In \textbf{Section} \ref{sec:Preprocessing}, we present a step-by-step instruction on how to pre-process the tweets, which are often tough to handle, and how we pick a collection of stopwords relevant to this research. This is followed by \textbf{Section} \ref{sec:Short Text Topic Clustering}, which provides a brief literature analysis in the subject of short text topic clustering and explains why this specific method was chosen above others. \textbf{Section} \ref{sec:GSDMM Algorithm} describes the procedures through which the unsupervised machine learning system monitors the spread of important narratives. In \textbf{Section} \ref{sec:Sentiment Analysis }, we demonstrated how we used the most recent advancements in NLP to do sentiment analysis on the tweets. 

Our findings are presented in \textbf{Section} \ref{sec:Results} together with the full procedure how to generate the narrative time series. In the last section of this work, we recommend options for further investigation.

\section{Data Collection and associate methods}
\label{sec:Data Collection}
We obtained >10M English-only tweets from Twitter API v2 Academic Research product subcategory for this investigation. For gathering the tweets, we collected tweets those with the keywords, “\textbf{cryptocurrencies}” and “\textbf{bitcoin}”. Due to the massive amount and our limited computing capabilities, we did not scrape the whole of the 2014–2021-time span. Instead, we used structural break analysis to the Bitcoin price data, which identifies sudden shifts in a time series, and found 9 such events between 2014 and 2021 (excluding 5\% of data at either end). We scraped tweets with those keywords around one month of those break instances. Roughly for each time period we collected ~2M tweets.

\begin{figure}[htp]
    \centering
    \includegraphics[width=0.3\textwidth]{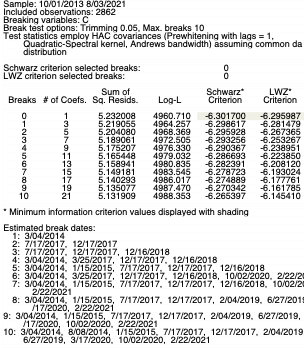}
    \caption{Structural break results and identified break instances}
    \label{fig:Structural break}
\end{figure}

\begin{table}[h]
\centering
\resizebox{0.5\columnwidth}{!}{%
\begin{tabular}{|c|c|}
\hline
\textbf{\begin{tabular}[c]{@{}c@{}}Date Indicating \\ the Structural Break\end{tabular}} &
  \textbf{\begin{tabular}[c]{@{}c@{}}Time Period around the structural \\ break date used to collect the tweets\end{tabular}} \\ \hline
4th March 2014     & 17/02/2014 to 19/03/2014 \\ \hline
8th August 2014    & 24/07/2014 to 23/08/2014 \\ \hline
15th January 2015  & 01/01/2015 to 30/01/2015 \\ \hline
17th July 2017     & 02/07/2017 to 02/08/2017 \\ \hline
4th February 2019  & 20/01/2019 to 19/02/2019 \\ \hline
27th June 2019     & 12/06/2019 to 13/07/2019 \\ \hline
17th March 2020    & 02/03/2020 to 02/04/2019 \\ \hline
2nd October 2020   & 16/09/2020 to 17/10/2020 \\ \hline
22nd February 2021 & 07/02/2021 to 06/03/2021 \\ \hline
\end{tabular}%
}
\vspace{0.5cm}
\caption{Date received from structural break and the time periods we scraped the tweets}
\label{tab:my-table}
\end{table}

\section{Methodologies}
\label{sec:Methodologies}

\subsection{Preprocessing the collected Twitter Data}
\label{sec:Preprocessing}
We converted millions of tweets to .csv file. The format of the data sent to us via the \textit{Twitter API v2} has not been altered or withdrawn.
Raw tweets are extremely unstructured and include redundant and frequently tricky information, making preprocessing crucial. There is a great deal of information in a tweet that we may not need or desire, depending on our goal(s). 
\\

"\#PiNetwork is worldwide, and every country should act to build \#GCV to support $1 = \$314,159$ barter. Fight for the common \#dream!

\#Pi \#GCV \#KYC \#Blockchain \#Crypto \#Currency \#Cryptocurrency \#Mainnet \#OpenMainnet \#PiMall \#PiChainMall \#PiMallChaina \#WorldEconomy \#PiNetworkLive @piNetwork"

\vspace{0.3cm}

For instance, the hashtags, links, and @ handles mentioned above may not be essential for our approach to topic modelling, since they do not give useful context for determining the tweet's intrinsic themes. We utilised a variety of NLP methods to clean the tweets and eliminate any irrelevant “\textit{noise}” for our purpose(s).\\
\begin{itemize}
    \item Eliminated all duplicate tweets as well as all tweets with \textit{NA} values.
    \item Nonsensical symbols, letters, numbers, and punctuation have been eliminated since we are primarily interested in words.
    \item Using regular expressions, all links, users, hashtags, and audio/video tags have been eliminated (Regex).
    \item All tweets have been lower-cased and all single letters have been deleted.
    \item The tweets are then tokenized, which is the process of converting a document (a tweet) into individual words.
    \item We deleted all stopwords from the tokenized tweets, and in the next part, we elaborated on stopwords and the collecting procedure for them in our context.
    \item In addition, before executing the Topic modelling algorithm, we ran Stemming on the tokenized tweets. Stemming is a natural language processing approach that reduces the inflection of words to their root forms, hence facilitating the preprocessing of text, words, and documents in preparation for text normalisation.
\end{itemize}

\subsubsection{Stopwords and extensive stop word collections, TF-IDF}
\label{sec:Stopwords}
We essentially cleanse the data to make it more suitable with the algorithms used for natural language processing and topic modelling. The greatest obstacle in sentiment analysis is stopwords that yield irrelevant sentiment values and stopwords that change the sentiment score towards neutral since they do not represent a certain sentiment. We deleted stopwords to make the data more useful for future preprocessing and topic modelling.

\subsubsection{Stopwords}
A stop word is a frequently occurring keyword that a search engine has been programmed to disregard while indexing items for searching and retrieving them in response to a search query. ‘the’, ‘a’, ‘an’, ‘and’, ‘in’ are instances of stop words.
We do not want these phrases to consume excessive database store space or processing time. By keeping track of the words that you think to be stop words, we can simply delete them. A list of stopwords is contained in Python's NLTK (Natural Language Toolkit). Additionally, the corpus may include a large number of contextual stopwords. To eliminate these stopwords, we used two key strategies.
\begin{itemize}
    \item \textbf{By incorporating nonwords and pronouns}: Python's NTLK module contains 40 stopwords by default; we added extra stopwords such as pronouns and themes that appear in every word, such as bitcoin; because bitcoin will be present in practically every tweet, it is useless for sentiment analysis since it moves the sentiment score to neutral.
    \item \textbf{TF-IDF Strategy}: TF-IDF stands for term frequency and document frequency inverted. This approach determines the word count of a group of documents. Typically, each word is assigned a score indicating its significance to the tweet and corpus. Information retrieval and text mining applications use this technique regularly.
\end{itemize}

I'll illustrate using the words “\textit{This building is really tall}.” Because we are acquainted with the semantics of the words and the phrase, comprehension is straightforward. But how can a computer programme (such as Python) comprehend this sentence? Any computer language may interpret textual material more effectively when it is represented numerically. Therefore, we must vectorize the whole text in order to effectively represent it.

By vectorizing tweets, we may do a variety of tasks, such as discovering relevant articles, rating, grouping, etc. When doing a Google search, this procedure is used (now they are updated to newer transformer techniques). The search string you enter is known as a question, while the web pages are known as tweets. The search engine consistently represents each and every tweet. When you do a search using a query, the search engine calculates the relevance of the query for each document, sorts them by relevance, and presents the top k documents. Throughout the procedure, the vectorized version of the question and the tweets are used.
\begin{multline}\\
    \text{t = term (word), d= tweets (set of words),}\\
    \\
    \text{N = count of corpus, Corpus= the total tweet set}\\
    \\
    \text{tf(t,d)} = \frac{\text{count of t in d}}{\text{number of words in d}}\\
    \\
    \text{df(t) = occurrence of t in N tweets}\\
    \\
    \text{IDF(t)} = \frac{\text{N}}{\text{df}}\\
\end{multline}

There are a few other issues with the IDF, such as the IDF value exploding for huge corpus sizes, like $N=10000$. Therefore, we use the IDF log to lessen the effect.
When a word is not in the vocabulary at inquiry time, it will simply be disregarded. But occasionally, when we employ a fixed vocabulary, and only a few vocabulary terms are present in the text, the pdf will be 0. We smooth the value by adding 1 to the denominator because we cannot divide by 0.
\begin{multline}\\
    \text{IDF(t)} = log(\frac{\text{N}}{\text{df+1}})\\
    \\
    \text{tf-IDF(t,d)} = \text{tf(t,d)} \times log(\frac{\text{N}}{\text{df+1}})\\
\end{multline}
The greater the TF-IDF score, the more significant the words in the tweet and the less common word. If the frequency of a word rises, so does its df value, and the IDF will fall logarithmically, resulting in a reduction in the TF-IDF score. The statistical information TF-IDF is designed to quantify the significance of a word within a collection (or corpus) of texts. IDF and TF-IDF are both 0 for these exceedingly frequent terms, which are mostly stopwords.
We picked TF-IDF over Bag of Words (BoW) for creating stopwords since Bag of Words (BoW) needs constructing a bag of comparable terms that may match the words in Tweets. Such a collection of terms cannot be developed for a dataset containing no conventional words, such as tweets.
BoW provides greater weight to the more frequent words and determines the meaning based on them; however, the more frequent words are often stopwords.

\subsection{Literature review of Short Text Topic Clustering }
\label{sec:Short Text Topic Clustering}
Short text clustering has always been a tough topic for many reasons:
\begin{itemize}
    \item Topic Modelling Using TF-IDF has less relevance since sentences include fewer words than passages.
    \item When dealing with high-dimensional data, using vector space leads in sparsity. This results in a high computation and memory storage need.
\end{itemize}

As a consequence of issues 1 and 2, calculating the number of clusters to split the text becomes more difficult, and we lose the capacity to parse our data into subjects.
Topic models are extensively used to extract latent themes in texts; however, they have limitations when dealing with sparse, high-dimensional, and large-volume data. Our primary emphasis here is Twitter, which generates relatively brief messages, and with the rising popularity of social media platforms like as Twitter, Google+, and Facebook, extracting latent ideas from short tweets has become more important.

\begin{figure}[htp]
    \centering
    \includegraphics[width=0.7\textwidth]{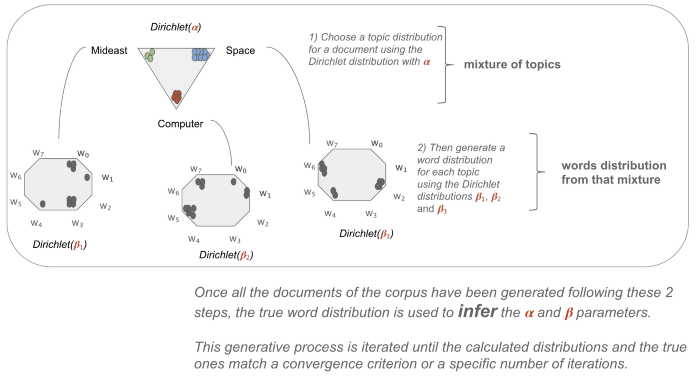}
    \caption{GSDMM Architecture}
    \label{fig:gsdmm}
\end{figure}

\begin{chapquote}{}
``\textbf{A document is generated by sampling a mixture of these topics and then sampling words from that mixture.}''
\end{chapquote}

However, research suggests that the most common topic model LDA \cite{blei2001latent} may not perform well when dealing with short and sparse text data, such as tweets, because these tend to focus on a particular topic, hence undermining the validity of the LDA's fundamental assumption \cite{mazarura2016comparison}, \cite{blei2001latent}. While LDA works well with larger texts (more than 50 words), it struggles with shorter ones. As a result, we used The Gibbs Sampling Dirichlet Mixture Model (GSDMM), a modified LDA approach that performs well on short text topic modelling tasks and assumes just one topic per document. Unlike the original LDA, all of the words inside a document are produced from the same distinct topic.
The LDA is a generative technique that believes each document in a corpus is created by a combination of themes \cite{yin2014dirichlet}. To produce document-topic and word-topic distributions, a Dirichlet distribution is used as a prior. The graphic below explains LDA's document production process utilising Dirichlet distribution in detail.

\subsection{GSDMM Algorithm}
\label{sec:GSDMM Algorithm}

To create cryptocurrency-related tales, I use the machine learning technique GSDMM, which is a modified version of the LDA algorithm. GSDMM, developed by Yin and Wang \cite{yin2014dirichlet}, is a Dirichlet Multinomial Mixture Model-based approach for short text clustering that labels textual documents in a manner similar to LDA, but with the key difference that the words within a document are generated using the same unique topic, rather than a mixture of topics.\\
According to Yin [8], the GSDMM has the following great qualities:
\begin{itemize}
    \item GSDMM can automatically identify the number of clusters.
    \item GSDMM provides a mechanism for balancing clustering results' completeness and homogeneity.
    \item The GSDMM converges very quickly.
    \item Unlike VSM-based approaches, GSDMM can deal with the sparse and high-dimensional problem of short texts.
    \item GSDMM, like other Topic Models, may select representative words for each cluster.
\end{itemize}

GSDMM is capable of dealing with the sparse and high-dimensional problem of short texts, as well as identifying the representative words for each cluster. GSDMM can also automatically estimate the number of clusters with a good balance of the completeness and homogeneity of the clustering discoveries and has a fast convergence rate.
GSDMM has a comparable technique called the Movie Group Process that will help us understand the many phases and processes behind the hood of STTM, as well as how to successfully adjust its hyper-parameters.
Consider a group of students who are seated at random at K tables in a restaurant (Clusters). They are all told to write down their favourite films (tweets) on a piece of paper (but it must remain a short list). The goal is to arrange them together so that students in the same group have similar movie tastes. To do so, pupils must choose a new table for each of the two following rules:
\begin{itemize}
    \item \textbf{Rule 1}: Choose a table with more students. By putting all students who are interested in the same movie to the same table, this rule improves completeness.
    \item \textbf{Rule 2}: Choose a table where students are interested in similar films. This rule aims to increase homogeneity, we want only members sharing the same movie’s interest at a table.
\end{itemize}

For the first portion of this model, it is essential to comprehend the Dirichlet distribution. The Dirichlet distribution is basically a multidimensional distribution. A distribution is just a probability distribution that represents the previous state likelihood of a document entering a cluster and the document's similarity to the cluster. Two factors ($\alpha$ and $\beta$ described given below) determine the form of the distribution.

After repeating this procedure, we anticipate that some tables will go and others will expand, resulting in clusters of students whose interests align with those of their movie. Simply put, this is what the GSDMM algorithm accomplishes!

\begin{figure}[htp]
    \centering
    \includegraphics[width=0.4\textwidth]{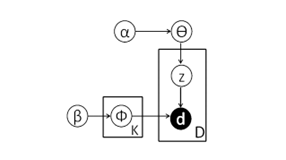}
    \caption{Graphical Model of DMM}
    \label{fig:gsdmm}
\end{figure}
This model's most important parameters are:\\

$\boldsymbol{\alpha}$: As previously explained, $\alpha$ is a parameter that influences the form of our probability distribution. In addition, $\alpha$ is calculated from the likelihood that a document will be sorted into a cluster. In the film example, this is the likelihood of a student selecting a table.\\

$\boldsymbol{\beta}$:  is the additional shape parameter for our distribution. $\beta$ is based on the similarity between the words in one text and those in another document. In relation to the movie groups, $\beta$ is the likelihood that a student will join a table with comparable movie preferences. If $\beta$ equals zero, for instance, the student will only connect tables containing movies in common. This may not be the optimal approach. It is possible that two students like thriller films, but they have not listed the same films. We still want kids who like suspense to end up in the same group.\\

$\boldsymbol{\phi}$: Using k clusters (mixtures), $\phi$ represents the multinomial distribution of clusters over words such that $p(w \mid z=k)=\varphi$ where $w=$ words and $z=$ cluster label.\\

$\boldsymbol{\theta}$: Similarly, $\theta$ is a multinomial distribution that accounts for $\alpha$, $p(d \mid z=$ $k)=\theta$, where $d=$ document.\\

These parameters culminate in the likelihood that a document $(d)$ is created by a cluster ( $k$ ) when Dirichlet priors are assumed.
$$
p(d \mid z=k)=\prod_{w \in d} p(w \mid z=k)
$$
Moreover, it is essential to note that the research assumes symmetric Dirichlet priors. This indicates that identical alphas and betas are expected from the outset. Alpha signifies that identical clusters are of equal importance, while beta suggests that identical words are of equal importance. Yin and Wang propose that in future rounds of this system betas should not have symmetric priors, and instead more popular words should have less relevance. If a term appears in every page, it is not a very useful indicator (Pitfalls of TF-IDF).

\subsubsection{GSDMM Results}
We run the GSDMM algorithm for all collected and preprocessed tweets. In this way, we reveal 4-5 topics/ narrative from the corpus for each of the specific time period.  The tables below show all the narratives revealed by the GSDMM for each specific time instances.

In each table “\textit{Top GSDMM words}” is the most representative words for each narrative display by the GSDMM algorithm, Sentiment score of each narrative is defined by Sentiment(S). Individual narratives are grouped and labelled into broader categories if needed for the interpretability denoted by “\textit{Labels (if any)}”.

\subsection{Sentiment Analysis}
\label{sec:Sentiment Analysis }
We are primarily interested in how social media narratives may truly explain the price changes of underlying crypto assets. We sought to achieve our goal by integrating two Text Mining approaches, Topic Modelling and Sentiment Analysis.

In accordance with Gavaldon and Larsen \cite{azqueta2020causal}, \cite{larsen2019business} I build each narrative-sentiment time series in a few simple steps with substantial process variation. Gavaldon \cite{azqueta2020causal} determined the sentiment of news articles using TextBlob Loria \cite{loria2018textblob} a lexicon and rule-based framework for Sentiment Analysis in English. We calculated the sentiment score for each tweet using pysentimiento.

Google's BERT \cite{devlin2018bert} was one of numerous ground-breaking advancements in NLP in recent years. Bidirectional Encoder Representation for Transformer (BERT) is a natural language processing (NLP) model developed by Google Research in 2018, and it has achieved cutting-edge accuracy on a variety of NLP tasks since its inception.

Encoder-decoder architecture refers to the encoder and decoder stacks of transformer architecture \cite{vaswani2017attention}, whereas BERT is only the transformer architecture's encoder stack. The architectural complexity of the BERT-base and BERT-large versions vary.
\begin{figure}[htp]
    \centering
    \includegraphics[width=0.6\textwidth]{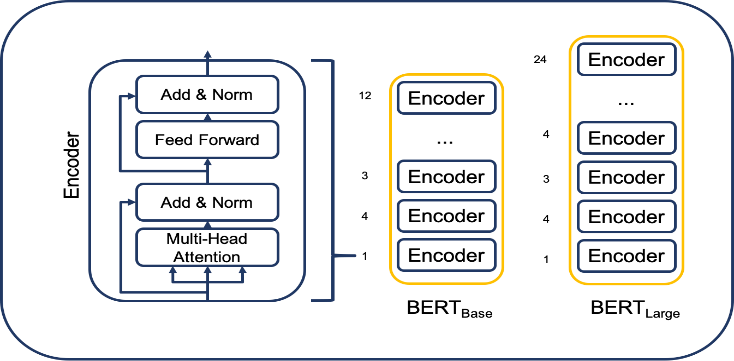}
    \caption{BERT transformer architecture}
    \label{fig:bert transformer architecture}
\end{figure}
Pysentimiento \cite{perez2021pysentimiento} is a multilingual Python toolkit trained on the SemEval 2017 dataset for Sentiment Analysis and other Social NLP applications (about 40k tweets). The basic model is BERTweet \cite{nguyen2020bertweet}, a RoBERTa \cite{liu2019roberta} model trained on English tweets with the same architecture as BERTbase. To create narrative-sentiment time series, we chose Pysentimiento over TEXTBLOB.\\

\begin{chapquote}{Robert J. Shiller}
``More serious attempts should be made to gather further narrative-sentiment time series data, moving beyond the passive collecting of others' words and toward trials that reveal meaning and psychological relevance.''
\end{chapquote}

Improving sentiment scores is one way to obtain better narrative-sentiment time series data. We employed BERTweet, which is taught using the RoBERTa \cite{liu2019roberta} pre-training technique, in addition to the traditional lexicon approach. RoBERTa improves performance robustness by optimising the BERT pre-training approach. The BERTweet pre-training dataset contains 850M English Tweets (16B word tokens 80GB), containing 845M Tweets streaming from 01/2012 to 08/2019 and 5M Tweets related to the COVID-19 outbreak. Experiments done by Nguyen \cite{nguyen2020bertweet} show that BERTweet outperforms the strong baselines RoBERTabase and XLM-Rbase \cite{conneau2019unsupervised} on the following three Tweet NLP tasks: text classification, named entity recognition, and part-of-speech tagging. Furthermore, according to P'erez \cite{perez2021pysentimiento}, the best-performing model in English is BERTweet, with a little advantage over RoBERTa. Both models are known to outperform BERT, with BERTweet being particularly well-suited to specific SocialNLP tasks due to its training solely on tweets. These findings push us to exploit BERTweet’s enormous potential for sentiment analysis and the Pysentimiento toolbox.
Pysentimiento employs the labels POS, NEG, and NEU to determine the sentiment of each tweet. In addition to labels, this toolbox provides sentiment scores for each label, which sum to 1. As an example, consider the following:

\begin{chapquote}{}
``Another merchant spotted in China accepting 100\% PiPayment! Pi Adoption is getting bigger everyday. Very Soon we will able to buy whatever we want at our doorstep and everywhere we go.
\#PiPayment \#PiNetwork \#Pioneers \#Bitcoin \#Cryptocurency \#LUNAC \#Cryptos Mainnet Binance
''
\end{chapquote}

The sentiment out of this tweet was,

\textit{“AnalyzerOutput(output=POS, probas={NEU: 0.05,NEG: 0.01,POS: 0.944})}”

\begin{figure}[htp]
  \centering
  \begin{minipage}[b]{0.5\textwidth}
    \includegraphics[width=\textwidth]{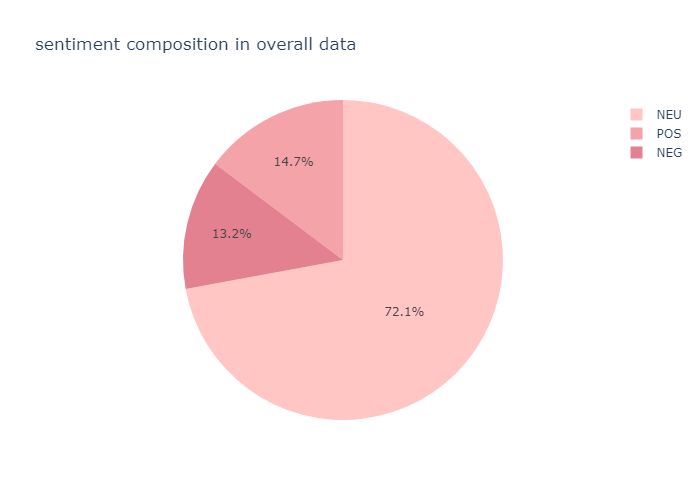}
  \end{minipage}
  \hfill
  \begin{minipage}[b]{0.5\textwidth}
    \includegraphics[width=\textwidth]{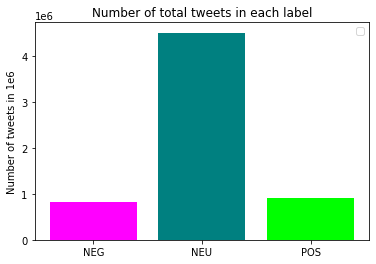}
  \end{minipage}
  \caption{Sentiment Composition for all the tweets}
\end{figure}

The accompanying graphics demonstrate the sentiment composition of all the tweets we gathered after applying the pysentimiento toolbox to the whole data. The same percentages of positive and negative labels reflect the fact that we scraped the data evenly around all structural break points where a rapid trend reversal would be seen. If the market exhibited bullish sentiment before to the structural break points, it will turn to bearish sentiment subsequently, and vice versa.

We provide our scaling method, which allows us to calculate a composite score $C$ for each tweet based on its POS, NEU, and NEG scores, where $C \in [-1, 1]$. We will utilise this composite score to create narrative-sentence time series in the future.

$$
C = Pos-Neg +(F(Neu) \times (Pos-Neg))
$$
Where $F$ is a function,

When a tweet sentiment is calculated using Roberta/BERTweet, the reasoning behind this method of calculating the composite score is that when the neutral score is the highest among the sentiment scores and there is a clear significant difference between the positive and negative scores, the neutral score will help the composite score to be biased towards the significant sentiment. When there is a small gap between positive and negative scores, the third component in the preceding equation tends to zero, regardless of the neutral sentiment's value. When there is a discernible bias in a specific tweet, the neutral value will be altered to bolster the dominating sentiment. Using different functions on the neutral score, the level of biassing and the extent to which the neutral score would affect the final composite score may be modified. First, we applied no function to the neutral score, and then we applied a square root function (as the value of the scores is between zero and one squaring will further reduce the value). We call the square root function as \textbf{cs2} and with our trial and error we found out this is working better than the no function and we adopt this for all our study.

\section{Results}
\label{sec:Results}
Tweets are far more continuous data streams than newspaper articles; we can verify that numerous tweets were sent in a single second. Such granularity cannot be correlated with crypto asset pricing. To acquire sentiment time series data for each narrative class, we aggregated all tweets of a given day from that narrative(s), and then calculated the sentiment score for that day by averaging the composite scores of all tweets from that day. 

\subsubsection{17 Feb to 19 Mar 2014}
\begin{figure}[htp]
    \centering
    \includegraphics[width=\textwidth]{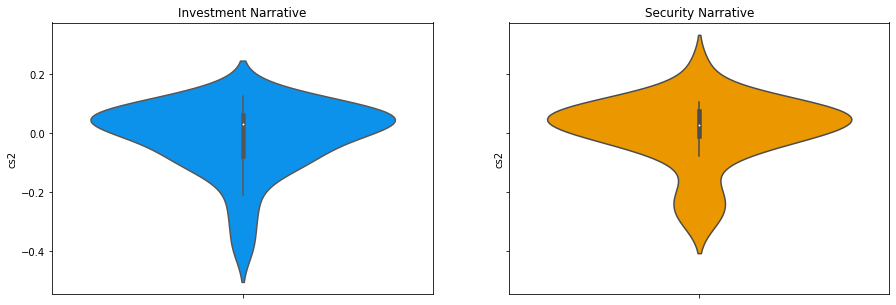}
    \caption{Violin Plot}
\end{figure}

\begin{figure}[htp]
    \centering
    \includegraphics[width=\textwidth]{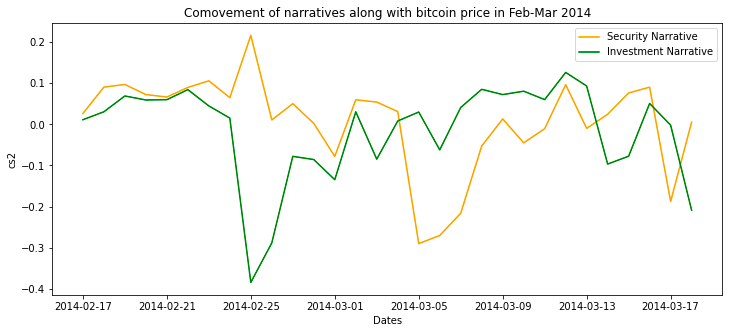}
    \caption{Co-Movement of Security Narrative and Investment Narrative}
\end{figure}
Mt. Gox was a bitcoin exchange with a Shibuya, Tokyo location. On February 24, 2014, Mt. Gox suspended trading, shut down its website and exchange company, and filed for bankruptcy protection against its creditors. Prior to this event, their CEO had resigned on February 23, 2014, and since February 7, 2014, tens of thousands of customers have been unable to withdraw cash from Mt Gox. Due to a series of these events between February 1, 2014 and the end of March, during the period of Mt. Gox's issues, the value of bitcoin plummeted by 36 percent.

Clearly, our data indicate that consumers were hesitant to engage in cryptocurrencies during this time period and had more security worries.  Investment and Security have an inverse relationship, as seen by a dramatic decline in investment narratives and a rapid increase in security narratives around February 24, 2014, when Mt. Gox ceased trading. The violin plot indicates that investment sentiment was more negative during this time period, which is corroborated by the fact that the average sentiment score for investment narratives was -0.014963.

\subsubsection{24 July to 23 August 2014}
\begin{figure}[htp]
    \centering
    \includegraphics[width=\textwidth]{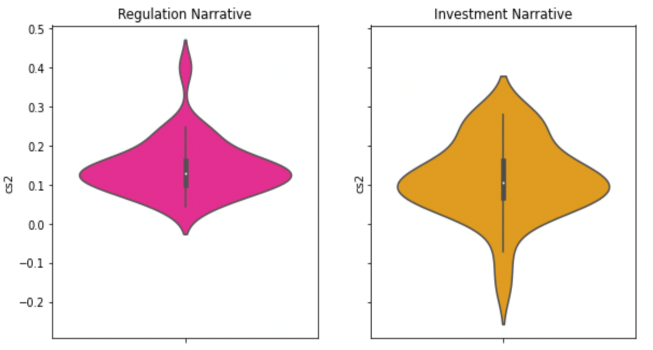}
    \caption{Violin Plot}
\end{figure}

\begin{figure}[htp]
    \centering
    \includegraphics[width=\textwidth]{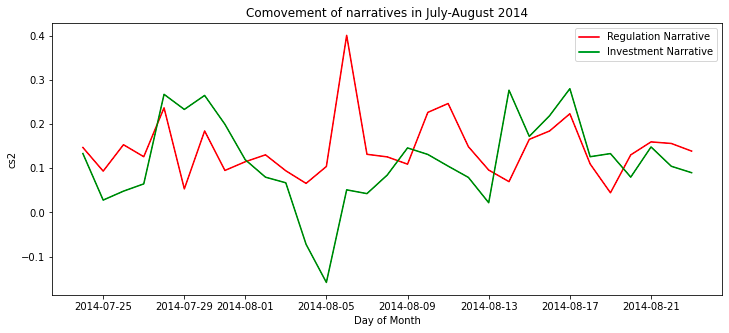}
    \caption{Co-Movement of Regulation Narrative and Investment Narrative}
\end{figure}

On August 4, 2014, New York State became the first state to regulate Bitcoin, causing a significant increase in sentiment towards regulation. As Bitcoin was meant to be unregulated and decentralized by default this news on regulation had a significant negative influence on the investing narrative. People have likely not yet recovered from the Mt. Gox event, and this has dealt a significant hit to investing sentiment.

As the co-movement of narratives graph demonstrates, Regulation and Investment narratives have a fairly inverse relationship, and we can clearly identify August 4, 2014 as the triggering date, as around that time an enormous decline in investment narrative coincides with an enormous increase in regulation narrative, similar to the previous scenario.

\subsubsection{1st Jan to 30th Jan 2015}
\begin{figure}[htp]
    \centering
    \includegraphics[width=\textwidth]{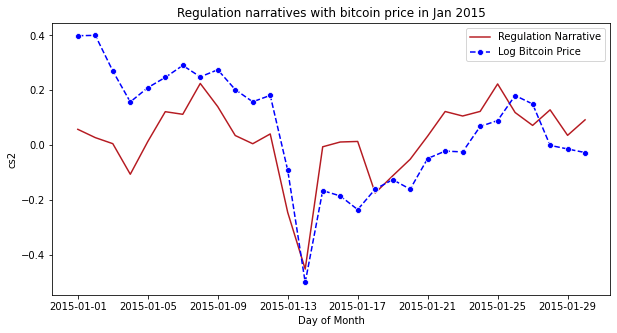}
    \caption{Co-Movement of Regulation Narrative and Log of Bitcoin Price}
\end{figure}
In the preceding graph, it is apparent that there is a significant relationship between the log of the bitcoin price and the narrative of regulation, and we can observe the massive drop in bitcoin price and sentiment owing to regulatory difficulties in Russia. During this time, Russia was on the edge of outlawing cryptocurrencies, beginning with a block on Bitcoin-related websites on the Internet. In addition, CEX.IO, a cloud mining company, has temporarily ceased operations due to the declining Bitcoin price.

\subsubsection{24 July to 23 August 2014}
\begin{figure}[htp]
    \centering
    \includegraphics[width=\textwidth]{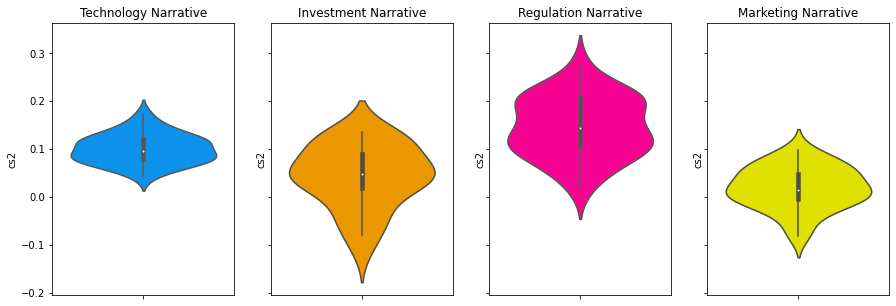}
    \caption{Violin Plot}
\end{figure}

\begin{figure}[htp]
    \centering
    \includegraphics[width=\textwidth]{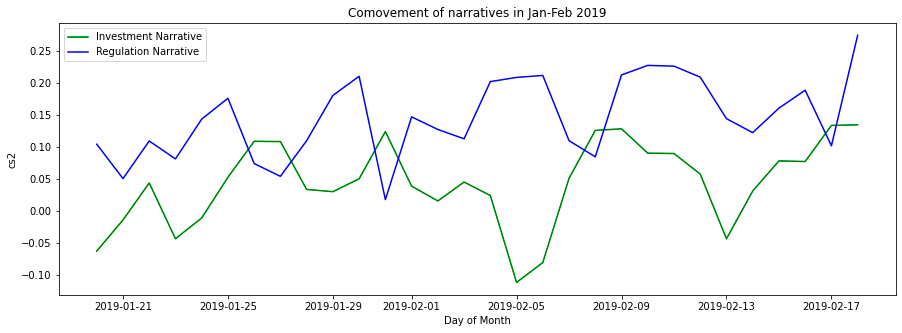}
    \caption{Co-Movement of Regulation Narrative and Investment Narrative}
\end{figure}

During this period, a series of events prompted widespread pessimism among investors, many of which were regulatory events, which explains why investing and regulatory narratives have an inverse association. The violin plot of many narratives bolsters the earlier co-movements of investment and regulatory narratives.
These are a few of the events that have contributed to the volatility of the cryptocurrency market and investor mood. 

\begin{itemize}
    \item According to the website of the state legislature, on January 31, the state of Wyoming will enable cryptocurrencies to be recognized as legal tender. The proposed legislation recognizes cryptocurrencies as intangible personal property, giving them the same status as traditional currency. Additionally, it permits banks to offer custody services for digital assets. The law will take effect on March 1.
    \item Argentinians are now able to load their public transportation cards with Bitcoin. This service is accessible in 37 sites nationwide. Prior to the advent of bitcoin payments, only PayPal was accepted.
    \item LocalBitcoins has declared that it will comply with the forthcoming AML and KYC rules of the European Union. In March, stricter user verification methods are anticipated to be released.
    \item It was revealed that an Argentine-Paraguayan commercial contract was settled in Bitcoin. Bitcoin was used to purchase \$7,100 worth of fumigation and chemical supplies from Argentina by Paraguay. The transaction establishes an intriguing precedent for international commerce, despite its modest size.
\end{itemize}

\subsubsection{2nd Mar to 2nd Apr 2020}
\begin{figure}[htp]
    \centering
    \includegraphics[width=\textwidth]{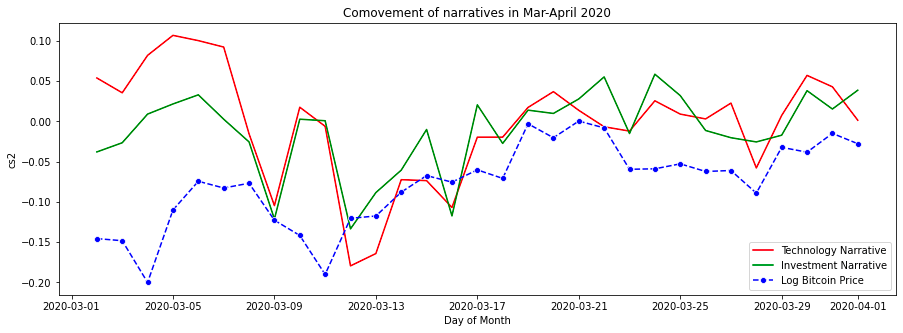}
    \caption{Co-Movement of Investment Narrative, Technology Narrative and Log of Bitcoin Price}
\end{figure}
During this period, after the covid plums, people gradually became interested in alternative investments, various governments pumped a great deal of money into the economy to stabilize it, and new generation investors were extremely optimistic about this asset, particularly the technology that comes with crypto. After this time, a remarkable boom of cryptocurrencies began, yielding returns of 1000\% in only one year. Despite this criticism, the advent of cryptocurrencies has exposed humanity to a new kind of technology known as blockchain. We see a substantial link between investment and technology narratives, as well as a correlation between the technology narrative and the bitcoin price log.

\subsubsection{2nd Mar to 2nd Apr 2020}
\begin{figure}[htp]
    \centering
    \includegraphics[width=\textwidth]{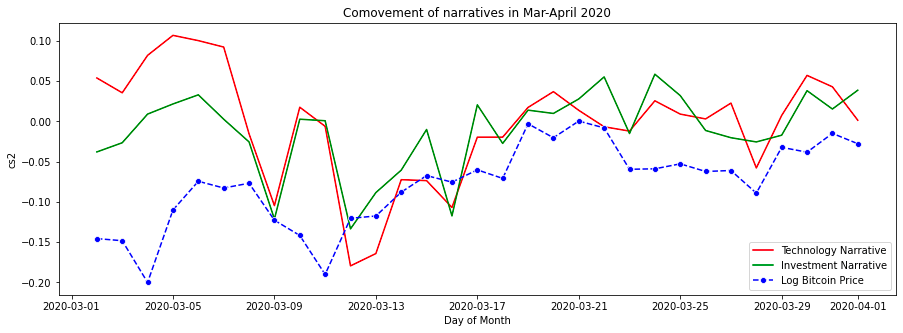}
    \caption{Co-Movement of Investment Narrative, Technology Narrative and Log of Bitcoin Price}
\end{figure}
During this period, after the covid plums, people gradually became interested in alternative investments, various governments pumped a great deal of money into the economy to stabilize it, and new generation investors were extremely optimistic about this asset, particularly the technology that comes with crypto. After this time, a remarkable boom of cryptocurrencies began, yielding returns of 1000\% in only one year. Despite this criticism, the advent of cryptocurrencies has exposed humanity to a new kind of technology known as blockchain. We see a substantial link between investment and technology narratives, as well as a correlation between the technology narrative and the bitcoin price log.

\subsubsection{16th Sep to 17th Oct 2020}
\begin{figure}[htp]
    \centering
    \includegraphics[width=\textwidth]{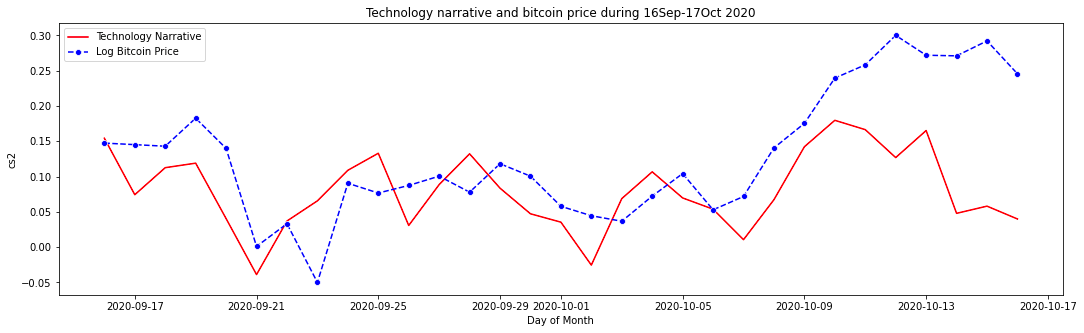}
    \caption{Co-Movement of Technology Narrative and Log of Bitcoin Price}
\end{figure}
Similar to the previous case, a substantial correlation exists between the Technology narrative and the log of bitcoin price.

\subsubsection{7th Feb to 6th Mar 2021}
\begin{figure}[htp]
    \centering
    \includegraphics[width=\textwidth]{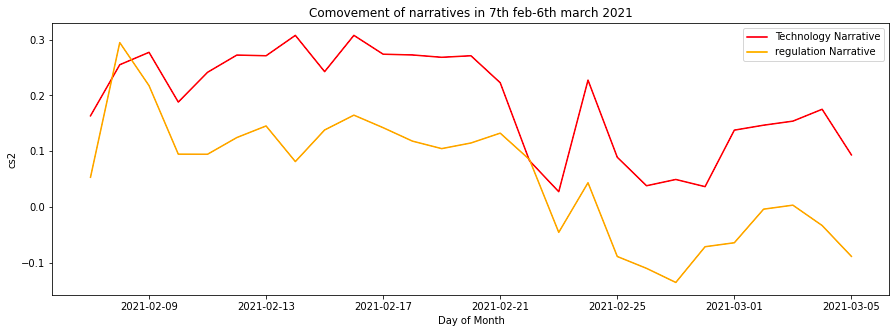}
    \caption{Co-Movement of Regulation Narrative and Technology Narrative}
\end{figure}
During this time period, we discovered an intriguing correlation between the Regulatory narrative and the Technology narrative. By this time, governments had begun to recognize the massive popularity of cryptocurrencies and most of them were no longer going to completely ban them. Furthermore, some governments welcomed the technological innovation side of cryptocurrencies and collaborated with them on serious issues such as alternative energy (Solana) and digital transactions (ETH). This adoption of technology demonstrates the intriguing connections between two narratives.

\section{Conclusion and Future Research}
In this paper we explored mainly two things, how to obtain narratives from short texts using a new kind of short text topic modelling tool and how to create the narrative time series using the state-of-the art algorithm for sentiment analysis. These narrative time series can be used with price of different crypto assets inferring valuable insights. And with the detail step by step process how to obtain narratives time series from textual data, our method can be used for any field of study rather than crypto or finance. Through our research, we discovered that the number of social media marketing clusters increased significantly between 2014 and 2021. These clusters are often sponsored advertising and contributed to our study as noise, therefore we've eliminated them. Of conjunction with social media marketing, we may also see a considerable growth in narrative media coverage from 2014 to 2021. This indicates that as time goes, more individuals have become aware of crypto, and thus, its popularity has grown substantially which may imply that the values of crypto have also increased as a result of this attention. We also see a significant spike in technology narratives, which suggests that people are increasingly becoming more interested in the technical breakthrough that accompanied crypto. We also observed a remarkable correlation between Regulation narratives and the price of crypto assets.

The findings of this research indicate that there is a connection between different types of narratives and the value of cryptocurrencies. Beliefs and requirements are the driving forces behind people, as well as the connections that form between them; they, in turn, have an effect on economic actors.

\section*{Acknowledgments}
The present endeavor was conducted under the auspices of Professor Abhijeet Chandra, and we express our profound gratitude to VGSOM, IIT Kharagpur for their generous provision of resources. Additionally, we extend our heartfelt appreciation to the Twitter team for granting access to the Twitter API V2, which greatly facilitated the process of data collection. Lubdhak would also like to extend recognition to the Government of India, specifically the Department of Science and Technology, for their support in the form of the Inspire SHE Scholarship. Furthermore, Lubdhak extends sincere thanks to Dr. Andrés Azqueta-Gavaldón for his invaluable contributions and insightful discussions during the initial stages of this project.

\section*{Author contributions}
The principal concept of the project was conceptualized and formalized by LM. UR, AS, BG, and SP contributed to the acquisition and analysis of the data. AC served as a mentor, providing essential resources and guidance as needed. LM took responsibility for composing the primary manuscript text and orchestrating the entirety of the project. LM, AS, BG, and SP collaborated in producing the visual representations in the form of Figures. The manuscript underwent rigorous review by all authors.

\bibliographystyle{unsrt}  
\bibliography{references}

\end{document}